\documentclass[11pt,a4paper]{article}

\usepackage{jheppub}
\usepackage{amsfonts}
\usepackage{amssymb}
\usepackage{comment}
\usepackage{epsfig}
\usepackage{graphicx}
\usepackage{amsmath}
\usepackage{tabu}
\usepackage[utf8]{inputenc}
\usepackage{latexsym}
\usepackage{tabularx}
\usepackage[nottoc]{tocbibind}
\usepackage{subfigure}
\usepackage{multirow}
\usepackage{booktabs}
\usepackage{enumerate}
\usepackage{mathrsfs}
\usepackage[section]{placeins}

\newcommand{\bea}{\begin{eqnarray}}
\newcommand{\eea}{\end{eqnarray}}
\newcommand{\bi}{\begin{itemize}}
\newcommand{\ei}{\end{itemize}}
\newcommand{\ben}{\begin{enumerate}}
\newcommand{\een}{\end{enumerate}}
\newcommand{\be}{\begin{equation}}
\newcommand{\ee}{\end{equation}}
\newcommand{\ba}{\begin{align}}
\newcommand{\ea}{\end{align}}
\newcommand{\comments}[1]{}

\newcommand{\mc}{\mathcal}

\newcommand{\beqa}{\begin{eqnarray}}
\newcommand{\eeqa}{\end{eqnarray}}

\title{Non-thermal CMSSM with a 125 GeV Higgs}

\author[1]{Luis Aparicio,}
\author[1,2,3]{Michele Cicoli,}
\author[4]{Bhaskar Dutta,}
\author[5]{Sven Krippendorf,}
\author[6]{Anshuman Maharana,}
\author[2,3]{Francesco Muia,}
\author[1,7]{Fernando Quevedo}

\affiliation[1]{ICTP, Strada Costiera 11, Trieste 34014, Italy}
\affiliation[2]{Dipartimento di Fisica e Astronomia, Universit\`a di Bologna, \\ via Irnerio 46, 40126 Bologna, Italy}
\affiliation[3]{INFN, Sezione di Bologna, via Irnerio 46, 40126 Bologna, Italy}
\affiliation[4]{Department of Physics and Astronomy, Mitchell Institute for Fundamental Physics \\
and Astronomy, TAMU, College Station, TX 77843-4242, USA}
\affiliation[5]{Rudolf Peierls Centre for Theoretical Physics, University of Oxford, 1 Keble Road, Oxford, OX1 3NP, UK}
\affiliation[6]{Harish Chandra Research Institute, Chhatnag Road, Jhunsi, Allahabad, UP 211019, India}
\affiliation[7]{DAMTP, Centre for Mathematical Sciences, Wilberforce Road, Cambridge, CB3 0WA, UK.}

\emailAdd{laparici@ictp.it}
\emailAdd{dutta@physics.tamu.edu}
\emailAdd{mcicoli@ictp.it}
\emailAdd{sven.krippendorf@physics.ox.ac.uk}
\emailAdd{anshumanmaharana@hri.res}
\emailAdd{muia@bo.infn.it}
\emailAdd{f.quevedo@damtp.cam.ac.uk}

\abstract{We study the phenomenology of the CMSSM/mSUGRA with non-thermal neutralino dark matter. Besides the standard parameters of the CMSSM we include the reheating temperature as an extra parameter. Imposing radiative electroweak symmetry breaking with a Higgs mass around $125$ GeV and no dark matter overproduction, we contrast the scenario with different experimental bounds from colliders (LEP, LHC), cosmic microwave background (Planck), direct (LUX, XENON100, CDMS, IceCube) and indirect (Fermi) dark matter searches. The allowed parameter space is characterised by a Higgsino-like LSP with a mass around $300$ GeV. The observed dark matter abundance can be saturated for reheating temperatures around $2$ GeV while larger temperatures require extra non-neutralino dark matter candidates and extend the allowed parameter space. Sfermion and gluino masses are in the few TeV region. These scenarios can be achieved in string models of sequestered supersymmetry breaking which avoid cosmological moduli problems and are compatible with gauge coupling 
unification. Astrophysics and particle physics experiments will fully investigate this non-thermal scenario in the near future.}

\preprint{DAMTP-2015-6 \\
\phantom{b} \hfill{MI-TH-1503}}


\begin{document}

\maketitle

\section{Motivation for non-thermal dark matter}
\label{sec:motivation}

One of the main particle physics candidates for dark matter (DM) is a stable neutralino $\chi$ which emerges as the lightest supersymmetric particle (LSP) in several scenarios beyond the Standard Model (SM). The DM relic abundance is generically \emph{assumed} to be produced thermally by the following process:
the LSP is in a thermal bath in the early universe, subsequently drops out of thermal equilibrium and freezes-out at temperatures of order
$T_f \simeq m_\chi/20$ when DM annihilation becomes inefficient.

However, we have no direct observational evidence of the history of the universe before Big Bang Nucleosynthesis (BBN)
for temperatures above $T_{\rm BBN} \simeq 3$ MeV. There is therefore no reason to assume a very simple cosmological history characterised by just a single period of radiation dominance from the end of inflation until BBN. In fact, the presence of a period of matter domination between the end of inflation and BBN could completely change the final prediction for the DM relic density if the reheating temperature at the end of this period of matter dominance is below $T_f$~\cite{MR,Moroi:1999zb}.

This non-thermal picture emerges generically in UV theories like string theory due to the ubiquitous presence
of gravitationally coupled scalars~\cite{CMP, NTDM, Allahverdi:2013noa}.
During inflation these fields, called moduli, get a displacement from their minimum that is in general of order $M_P$~\cite{DRT}. After the end of inflation,
when the Hubble constant reaches their mass, $H\sim m_{\rm mod}$, they start oscillating around their minimum and store energy.
Redshifting as matter, they quickly dominate the energy density of the universe which gets reheated when the moduli decay.
Being only gravitationally coupled, the moduli tend to decay very late when $H\sim \Gamma \sim m_{\rm mod}^3/M_P^2$.
The corresponding reheating temperature
\be
T_R \sim \sqrt{\Gamma M_P}\sim m_{\rm mod} \sqrt{\frac{m_{\rm mod}}{M_P}}\,,
\ee
has to be larger than $T_{\rm BBN}$ in order to preserve the successful BBN predictions.\footnote{$T_R$ has also to be lower than the temperature above which the internal space decompactifies~\cite{TempDecomp}.}
This requirement sets a lower bound on the moduli masses of order $m_{\rm mod}\gtrsim 30$ TeV~\cite{CMP}.

Generically in string compactifications supersymmetry (SUSY) breaking effects develop a mass for the moduli
and generate by gravity or anomaly mediation soft terms of order $M_{\rm soft}$. Due to their common origin, the mass of the lightest modulus $m_{\rm mod}$
is therefore related to the scale of the soft terms as $M_{\rm soft} = \kappa m_{\rm mod}$.
Given the cosmological constraint $m_{\rm mod}\gtrsim 30$~TeV, only models with $\kappa\ll 1$ can allow for
low-energy SUSY to solve the hierarchy problem.
Values of $\kappa\sim\mc{O}(10^{-2})$ can come from loop suppression factors~\cite{Mirage,G2,Dudas:2012wi} while much smaller values $\kappa\sim\mc{O}(10^{-3}-10^{-4})$ can arise due to sequestering effects~\cite{seqLVS, Aparicio:2014wxa}. For $M_{\rm soft}\sim \mc{O}(1)$ TeV, the corresponding reheating temperature becomes
\be
T_R \sim  \frac{M_{\rm soft}}{\kappa^{3/2}} \sqrt{\frac{M_{\rm soft}}{M_P}} \sim \,\kappa^{-3/2}\,\mc{O}(10^{-2})\,\,{\rm MeV}\,,
\ee
which for $10^{-2}\lesssim\kappa\lesssim 10^{-4}$ is between $\mc{O}(10)$ MeV and $\mc{O}(10)$ GeV.
This is below the freeze-out temperature for LSP masses between $\mc{O}(100)$ GeV and $\mc{O}(1)$ TeV which is
$T_f \sim \mc{O}(10-100)$ GeV. Therefore any DM relic density previously produced via the standard thermal mechanism gets erased by the late-time decay of the lightest modulus. In this new scenario, the LSP gets produced non-thermally from the modulus decay.

From a bottom-up perspective, non-thermal cosmological histories can also enlarge the available parameter space of different DM models consistent with direct and indirect detection experiments, due to the presence of the additional parameter $T_R.$ This is appealing as it is very hard to reproduce a correct thermal relic density in the CMSSM/mSUGRA (see for instance~\cite{Baer:2012uya}) since a Bino-like LSP tends to overproduce DM
(apart from some fine-tuned cases like stau co-annihilation and A-funnel or in the case of precision gauge coupling unification~\cite{Krippendorf:2013dqa}) while for a Higgsino- or Wino-like LSP the relic density is in general underabundant (except for cases like well tempered Bino/Higgsino or Bino/Wino DM~\cite{ArkaniHamed:2006mb}).

The purpose of this paper is to study non-thermal DM in the CMSSM/mSUGRA where the free parameters are: the standard parameters of the CMSSM/mSUGRA~\cite{msugra},
i.e.~the universal scalar mass $m$, gaugino mass $M$ and trilinear coupling $A$ defined at the GUT scale, $\tan\beta$ and the sign of $\mu$, with in addition the reheating temperature $T_R$ from the decay of the lightest modulus. We shall follow the RG running of these parameters from the GUT to the electroweak (EW) scale and require a Higgs mass $m_h \simeq 125$ GeV, a correct radiative EW symmetry breaking (REWSB) and no DM non-thermal overproduction.
We shall then focus on the points satisfying all these requirements and we will impose on them several phenomenological constraints coming
from LEP~\cite{lep}, LHC~\cite{lhc}, Planck~\cite{planck}, Fermi (pass 8 limit)~\cite{fermi}, XENON100~\cite{xenon}, CDMS~\cite{cdms}, IceCube~\cite{icecube} and LUX~\cite{lux}. Moreover we shall focus only on cases where the LSP has a non-negligible Higgsino component since Bino-like DM requires a very low reheating temperature which is strongly disfavoured by dark radiation bounds in the context of many string models~\cite{Allahverdi:2014ppa}. Interestingly we shall find that the constraints from Fermi and LUX are very severe and do not rule out the entire non-thermal CMSSM parameter space only for reheating temperatures $T_R\gtrsim \mc{O}(1)$ GeV. The best case scenario is realised for $T_R = 2$ GeV where a Higgsino-like LSP with a mass around $300$ GeV can saturate the observed DM relic abundance. For larger reheating temperatures the LSP Bino component has to increase, resulting in strong direct detection bounds which allow only for cases with DM underproduction. Values of $T_R$ above $1$ GeV require 
values of $\kappa\sim\mc{O}(10^{-3}-10^{-4})$ which can be realised only in models where the CMSSM is sequestered from the sources of SUSY breaking~\cite{seqLVS, Aparicio:2014wxa}. Apart from DM, these models are very promising since they can be embedded in globally consistent Calabi-Yau compactifications~\cite{CYembedding}, allow for TeV-scale SUSY and successful inflationary models~\cite{KMI}, do not feature any cosmological moduli problem,\footnote{Ref.~\cite{Dutta2015x} provides a significant bound on moduli masses and the number of e-foldings during inflation which can be a challenge for many models.} are compatible with gauge coupling unification and do not suffer from any moduli-induced gravitino problem~\cite{gravProbl}.

In Sec.~2 we discuss CMSSM soft terms, in Sec.~3 we analyse the
non-thermal CMSSM, in Sec.~4 we discuss our results and conclusions are
given in Sec.~5.

\section{CMSSM soft terms}

In a UV completion of the MSSM like string theory, SUSY is spontaneously
broken by some dynamical mechanism which generates particular relations
between the soft terms via gravity, anomaly or gauge mediation. In the
case when the soft terms are universal at the GUT scale, they are given by
the scalar mass $m$, the gaugino mass $M$, the trilinear coupling $A$ and
the bilinear Higgs mixing $B$. We can generically parameterise these soft
terms and the $\mu$ parameter at the GUT scale as:
\be
m = a\ |M|\,,\quad A = b\ M\,,\quad B = c \ M\,, \quad \mu = d\ M\,,
\label{CMSSM}
\ee
where, in a stringy embedding, the coefficients $a$, $b$, $c$ and $d$ are
functions of the underlying parameters while the gaugino mass $M$ sets the
overall scale of the soft terms in terms of the gravitino mass $m_{3/2}$.
In order to perform a phenomenological analysis of this scenario one has
to follow the renormalisation group (RG) evolution of these soft terms
from the GUT to the EW scale and impose the following constraints: a
correct REWSB, a Higgs mass of order $m_h \simeq 125$ GeV, no DM
overproduction and no contradiction with flavour observables and with any
experimental result in either particle physics or cosmology.

A viable REWSB can be obtained if at the EW scale the following two
relations are satisfied:
\be
\mu^2 = \frac{m_{H_d}^2-m_{H_u}^2 
\tan^2\beta}{\tan^2\beta-1}-\frac{m_Z^2}{2}\,,
\label{resb1}
\ee
where:
\be
\sin(2\beta)=\frac{2 |B\mu|}{m_{H_d}^2+m_{H_u}^2 + 2\mu^2}\,.
\label{resb2}
\ee
Given that the requirement of a correct REWSB fixes only the magnitude of
$\mu$ leaving its sign as free, the parameters (\ref{CMSSM}) are typically
traded for the standard CMSSM/mSUGRA parameters:
\be
m = a\ |M|\,,\quad A = b\ M\,,\quad \tan\beta\equiv \frac{\langle
H_u^0\rangle}{\langle H_d^0\rangle}\,, \quad {\rm sign}(\mu)\,,
\label{CMSSMparameters}
\ee
where one runs $m$, $M$ and $A$ (or $a$, $b$ and $M$ in our case) from the
GUT to the EW scale
with a particular choice of $\tan\beta$ and sign$(\mu)$.
Eqs.~(\ref{resb1}) and (\ref{resb2}) then give the value of $B$ and $\mu$
at the EW scale. This is the way in which typical spectrum generators operate.\footnote{Here we use \texttt{SPheno} v3.3.3~\cite{spheno}.}
The boundary values of $B$ and $\mu$ at the GUT scale which give a correct
REWSB can be obtained by running back $B$ and $\mu$ from the EW to the GUT
scale. 
In this way we obtain the values of the coefficients $c$ and $d$.
In a viable UV model, these values of $c$ and $d$ have to be compatible
with the values allowed by the stringy dynamics responsible for SUSY
breaking and the generation of soft terms.

\section{Non-thermal CMSSM}
\label{SecResults}

As motivated in Sec.~\ref{sec:motivation}, we shall consider scenarios where the LSP is produced non-thermally like in the case of string compactifications where the reheating temperature $T_R$ from the decay of the lightest modulus is generically below the thermal freeze-out temperature~\cite{NTDM, Allahverdi:2013noa}.
This reheating temperature represents an additional parameter which has to be supplemented to the standard free parameters of the CMSSM ($a$, $b$, $M$, $\tan\beta$ and the sign of $\mu$). We call this new scenario the `non-thermal CMSSM' which is characterised by the following free parameters: $T_R$, $a$, $b$, the gaugino mass $M$, $\tan \beta$ and the sign of $\mu$.

\subsection{Non-thermal dark matter relic density}

The abundance of DM particles $\chi$ produced non-thermally by the decay of the lightest modulus is given by~\cite{Moroi:1999zb}:
\bea
\label{dmdens}
\left({n_\chi \over s}\right)^{\rm NT} = {\rm min} \left[\left({n_\chi \over s}\right)_{\rm obs}
{\langle \sigma_{\rm ann} v \rangle_f^{\rm Th} \over \langle \sigma_{\rm ann} v \rangle_f} \sqrt{\frac{g_*(T_f)}{g_*(T_R)}}\left({T_f \over T_R}\right),
Y_{\rm mod}~ {\rm Br}_\chi \right] \, , \nonumber \\
\,
\eea
where $g_*$ is the number of relativistic degrees of freedom, $\langle \sigma_{\rm ann} v \rangle_f^{\rm Th}\simeq 2 \times 10^{-26} {\rm cm}^3\,{\rm s}^{-1}$
is the annihilation rate at the time of freeze-out needed in the thermal case to reproduce the observed DM abundance:
\be
\left({n_\chi \over s}\right)_{\rm obs} = \left(\Omega_\chi h^2\right)_{\rm obs} \left(\frac{\rho_{\rm crit}}{m_\chi s h^2}\right) \simeq 0.12 \left(\frac{\rho_{\rm crit}}{m_\chi s h^2}\right),
\label{obs}
\ee
whereas the yield of particle abundance from modulus decay is:
\be
Y_{\rm mod} \equiv {3 T_R \over 4 m_{\rm mod}} \sim \sqrt{\frac{m_{\rm mod}}{M_P}}\,.
\label{yield}
\ee
${\rm Br}_\chi$ denotes the branching ratio for modulus decays into $R$-parity odd particles which subsequently decay to DM.

The expression (\ref{dmdens}) leads to two scenarios for non-thermal DM:
\ben
\item `Annihilation scenario': in this case the DM abundance is given by the first term on the right-hand side of eq.~(\ref{dmdens})
and the DM particles undergo some annihilation after their initial production by modulus decay.
In order to avoid DM overabundance one needs
\be
\langle \sigma_{\rm ann} v \rangle_f \geq \langle \sigma_{\rm ann} v \rangle_f^{\rm Th}\,\sqrt{\frac{g_*(T_f)}{g_*(T_R)}} \left(\frac{T_f}{T_R}\right).
\ee
Given that $T_R < T_f$ and $g_*(T_R) < g_*(T_f)$, this scenario requires
$\langle \sigma_{\rm ann} v \rangle_f > \langle \sigma_{\rm ann} v \rangle_f^{\rm Th}$ as in the case of thermal underproduction.
This condition is satisfied by a Higgsino- or Wino-like LSP but not by a pure Bino-like LSP which would generically lead to non-thermal overproduction (apart from the aforementioned cases). However, given that we shall focus on models with just gravity mediated SUSY breaking (contributions from anomaly mediation are subleading) and universal gaugino masses at the GUT scale as in~\cite{seqLVS, Aparicio:2014wxa}, the LSP can never be Wino-like due to the RG running of the gauginos.\footnote{Moreover a Wino-like LSP has a significantly larger annihilation cross section than a Higgsino-like LSP resulting in a strong conflict with Fermi bounds for sub-TeV Wino-like DM~\cite{Cohen:2013ama}.}
In this context, the `Annihilation scenario' requires a Higgsino-like DM. Let us finally point out that the non-thermal DM relic density can be written in terms of the thermal one as:
\be
\Omega_\chi^{\rm NT} h^2 = \sqrt{\frac{g_*(T_f)}{g_*(T_R)}}\left({T_f \over T_R}\right) \Omega_\chi^{\rm Th} h^2\,.
\label{NTdm}
\ee
For $5$ GeV $< T_f < 80$ GeV (corresponding to $100$ GeV $< m_\chi < 1.6$ TeV), the top, the Higgs, the Z and the W$^\pm$ are not relativistic, giving $g_*(T_f)=86.25$ and:
\be
\Omega_\chi^{\rm NT} h^2 = 0.142 \,\sqrt{\frac{10.75}{g_*(T_R)}}\left({m_\chi \over T_R}\right) \Omega_\chi^{\rm Th} h^2\,.
\label{nonthermaldm}
\ee

\item `Branching scenario': in this case the DM abundance is given by the second term on the right-hand side of eq.~(\ref{dmdens})
and the DM particles are produced directly from the modulus decay since their residual annihilation is inefficient.
In this case both large and small cross sections can satisfy the DM content since the annihilation cross section does not have any impact on the DM content. This scenario is very effective to understand the DM and baryon abundance coincidence problem~\cite{Allahverdi:2010im}.
Given that in general we have ${\rm Br}_\chi\gtrsim 10^{-3}$, in order not to overproduce DM for LSP masses of order hundreds of GeV, one needs $Y_{\rm mod}\lesssim 10^{-9}$. This condition requires a very low reheating temperature:
\be
T_R\lesssim 10^{-9} m_{\rm mod} = 10^{-9}\kappa^{-1} M_{\rm soft}\,.
\ee
For $M_{\rm soft}\sim \mc{O}(1)$ TeV and $\kappa\sim\mc{O}(10^{-2}-10^{-4})$ we find $T_R\lesssim \mc{O}(10)$ MeV. In order to obtain such a low reheating temperature one has in general to consider models where the modulus coupling to visible sector fields is loop suppressed~\cite{Allahverdi:2013noa}. However in this case it is very challenging to avoid a large modulus branching ratio into hidden sector light fields like stringy axions~\cite{DR} and so typically dark radiation is overproduced~\cite{Allahverdi:2014ppa}. Therefore the `Branching scenario' does not seem very promising from the phenomenological point of view.
\een

\subsection{Collider and CMB constraints}

\begin{figure}[!ht]
\centering
\includegraphics[width=7.5cm]{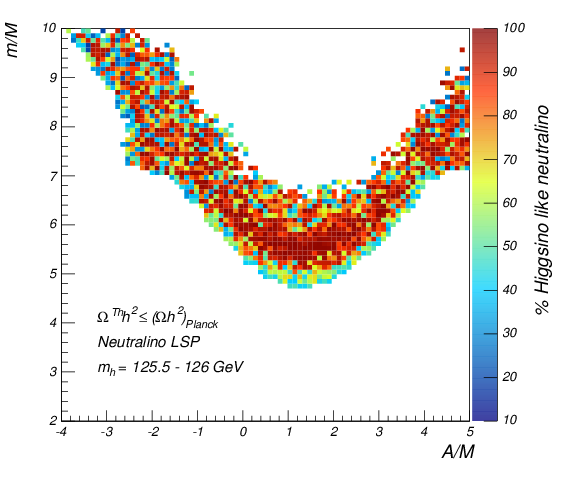}
\includegraphics[width=7.5cm]{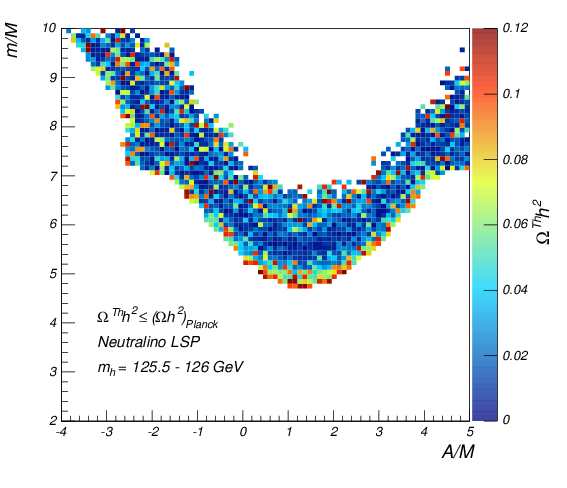}
\caption{Correlation between $a=m/M$ and $b=A/M$ for different LSP compositions (left) and DM thermal relic densities (right)
in the region where the LSP is at least 10\% Higgsino.}
\label{Fig3}
\end{figure}

\begin{figure}[!ht]
\centering
\includegraphics[width=7.5cm]{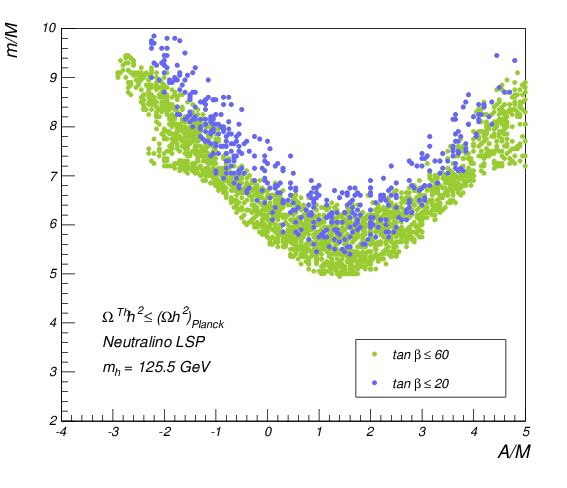}
\includegraphics[width=7.5cm]{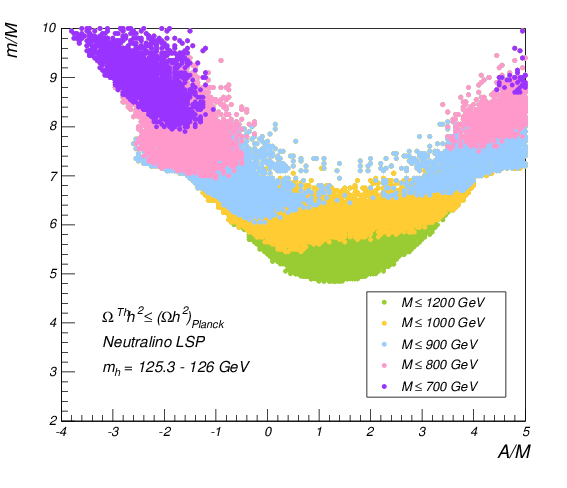}
\caption{Correlation between $a=m/M$ and $b=A/M$ for different values of $\tan\beta$ (left) and gaugino mass (right)
in the region where the LSP is at least 10\% Higgsino.}
\label{Fig4}
\end{figure}

\begin{figure}[!ht]
\centering
\includegraphics[width=7.5cm]{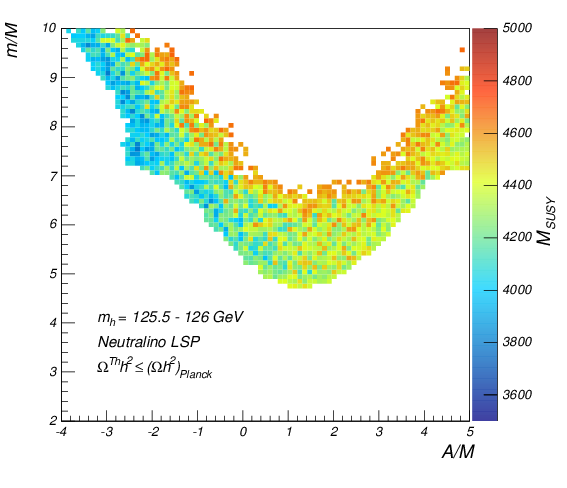}
\includegraphics[width=7.5cm]{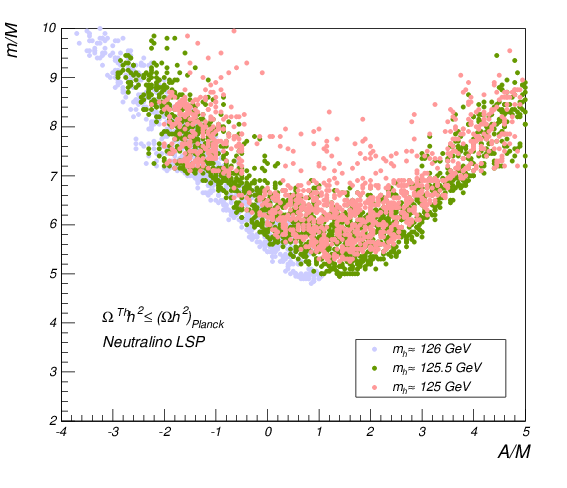}
\caption{Correlation between $a=m/M$ and $b=A/M$ for different values of the averaged stop mass $M_{\rm SUSY}$ (left) and the Higgs mass (right)
in the region where the LSP is at least 10\% Higgsino.}
\label{Fig5}
\end{figure}

Due to the considerations mentioned above, if the LSP is Bino-like we generically get DM overproduction also in the non-thermal case. We shall therefore look for particular regions in the non-thermal CMSSM parameter space where the LSP has a non-negligible Higgsino component. We have developed a Monte Carlo programme to find the regions of this parameter space where the LSP is Higgsino-like, the Higgs mass is around $125.5$ GeV,\footnote{Both ATLAS and CMS give values of the Higgs mass between $125$ and $126$ GeV. In what follows,  we will consider ranges of values in this region, allowing to some extent for the uncertainty in the spectrum generators as well. Allowing for a larger uncertainty in the Higgs mass does not alter the following results qualitatively.}
REWSB takes place correctly and the following phenomenological constraints are satisfied: 
\bi
\item LEP~\cite{lep} and LHC~\cite{lhc} constraints on neutralino and chargino direct production: $m_\chi \gtrsim 100$ GeV;
\item LHC~\cite{lhc} bounds on gluino and squark masses: $m_{\tilde{g}} \gtrsim 1300$ GeV;
\item LHC constraints from flavour physics: BR$(B_s \rightarrow \mu^+ \mu^-)$~\cite{lhcb} and the constraint on BR$(b\rightarrow s\gamma)$~\cite{pdg};
\item Planck data on DM relic density~\cite{planck}.
\ei
To avoid complications with the applicability of the standard \texttt{Spheno} version, we restrict our scan on the following parameter ranges: $\tan\beta=1$ to $55,$ $a=0\,{\rm to}\,10,$ $b=-5\,{\rm to}\,5,$ and the universal gaugino mass at the high scale $M=0.3-3$~TeV. The results are shown in Figs. \ref{Fig3}-\ref{Fig5} for positive $\mu$ (the LSP relic density has been calculated using \verb"micrOMEGAS" v3~\cite{micromegas}). 

The plots in Fig.~\ref{Fig3} show the points surviving the above constraints in the $A/M$-$m/M$ plane (at the GUT scale). The points fit into a V-shaped band illustrating a slight hierarchy between scalar and gaugino masses ($m\gtrsim 5 M$) and values of $A$ almost symmetric around $A\simeq M$. The regions shown in the plot are mostly for $T_R$ rather smaller than $T_f$ which keeps mostly the focus point regions in the allowed parameter space. The coannihilation and A-funnel regions can also contribute to the allowed parameter space but they are very fine tuned. The V-shape of our plots is caused by the focus point region which can be obtained by setting $\mu\sim m_Z$ in the EWSB condition with loop corrections. In fact, the dependence of $\mu^2$ on $A$, $M$ and $m$ arises through $m_{H_u}^2$ which depends on the UV soft terms in the following way: $M^2(f(Q) + g(Q) A/M+ h(Q) (A/M)^2 + e(Q) (m/M)^2)$, where $f$, $g$, $h$ and $e$ depend on dimensionless gauge and Yukawa couplings ($e$ also includes the tadpole correction from the stop loop)
 and $Q$ is the SUSY breaking scale~\cite{FMM}. A leading order cancellation in this expression, 
as needed to achieve a small $\mu$-term in (\ref{resb1}), gives a V-shaped band in the $A/M$-$m/M$ plane. We also apply the Higgs mass constraint in this parameter space which depends on the square of $X_t\equiv A_t-\mu \cot\beta$~\cite{Carena:1995bx} and $X_t^2$ preserves the V-shape due to its dependence on $(A/M)^2$.

To illustrate the allowed parameter range and to illustrate different aspects of the surviving points, we show the following plots (for positive sign of $\mu$):
\begin{enumerate}
\item In Fig.~\ref{Fig3}, the colour codes illustrate the percentage of Higgsino-like neutralino on the left plot and the neutralino contribution to the thermal DM relic density on the right. Note that in most of the points the neutralinos contribute only a small percentage of the total DM relic density and other DM candidates, such as axions, have to be present. The thermal DM relic density is close to the observed Planck value only in a small region corresponding to an LSP that is approximately 50\% Bino and 50\% Higgsino. However, as we will see in the next section, after imposing indirect detection constraints from Fermi, the only region which survives is the one where thermal DM is underabundant (by about 10\% of the observed relic density). On the other hand, as we shall see in the next section, non-thermal Higgsino-like DM can lead to larger relic densities which can saturate the Planck value for reheating temperatures around $2$ - $3$ GeV.

\item The colour codes in Fig.~\ref{Fig4} illustrate the dependence on $\tan\beta$ on the left plot and different values of gaugino masses on the right plot with well-defined domains for different ranges of gaugino masses inside the V-shaped band. Note that $\tan\beta$ tends to have larger values as expected from the fit of $m_h$ and the parameter ranges in this scan. Smaller values of $A/M$ and $m/M$ are preferred for larger gaugino mass due to RG flow of masses to fit the experimental value of $m_h$.

\item For Fig.~\ref{Fig5} the colours illustrate on the left the different values of the typical scale of SUSY particles $M_{\rm SUSY}$, defined here as the averaged stop mass $M_{\rm SUSY}^2=m_{\tilde{t}_1} m_{\tilde{t}_2}$. Notice that $M_{\rm SUSY}$ is around $4$ - $5$ TeV. In principle, we could explore values larger than $5$ TeV however it would bring us beyond the level of applicability of the spectrum generator \verb"SPheno" we have been using which assume similar values for all soft terms. An analysis for a split-like SUSY case with larger differences between sfermions and gaugino masses would be required in that case but this goes beyond the scope of this article. The colours on the right plot illustrate the dependence on the Higgs mass for which we have taken $m_h=125, 125.5, 126$ GeV respectively. Note that for $m_h=126$ GeV there are allowed points only on the left of the V-shaped band because of the above mentioned cut-off on $M_{\rm SUSY}$. Generally speaking we see that by allowing a larger range for the Higgs mass, we widen the V-shaped region. 
\end{enumerate}

\subsection{Direct and indirect detection constraints}

In the figures above we have set $\mu>0$ but their pattern does not change for $\mu<0$. The next step is to impose the following phenomenological constraints for the separate case of positive and negative $\mu$ since the DM direct detection cross section depends on sign$(\mu)$:
\bi
\item Fermi bounds on DM indirect detection~\cite{fermi};
\item IceCube~\cite{icecube} and XENON100~\cite{xenon} bounds for spin dependent DM direct detection;
\item LUX~\cite{lux}, CDMS~\cite{cdms} bounds on spin independent DM direct detection.
\ei

\subsubsection{Results for positive $\mu$}

If we impose the above constraints, indirect detection bounds turn out to be very severe. 
Using the new Fermi bounds (pass 8 limit)
coming from data collected until 2014 (the pass 7 limit includes only data until 2012), we do not find any allowed point for $T_R\lesssim 2$ GeV.

\begin{figure}[!ht]
\centering
\includegraphics[width=7.5cm]{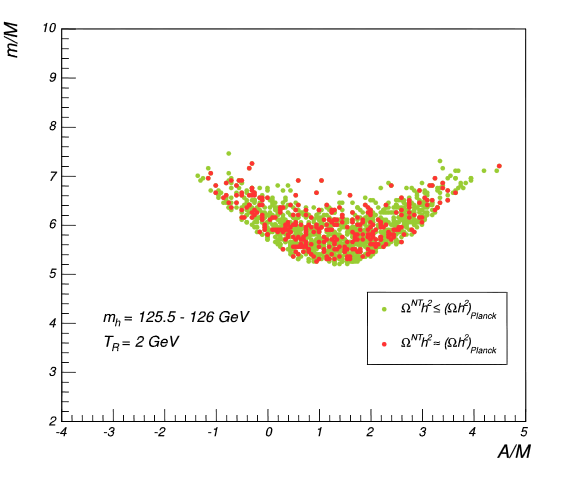}
\includegraphics[width=7.5cm]{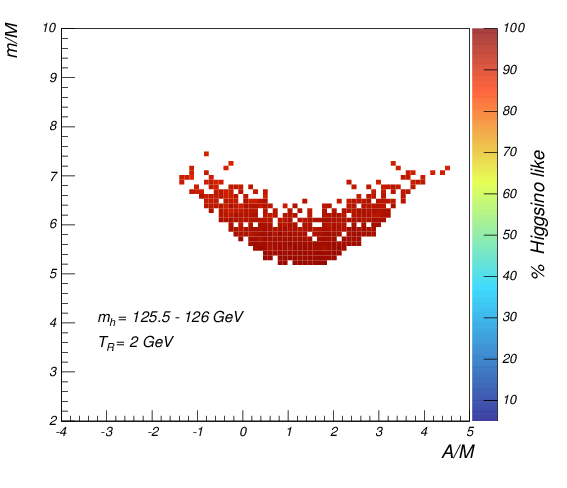}
\includegraphics[width=7.5cm]{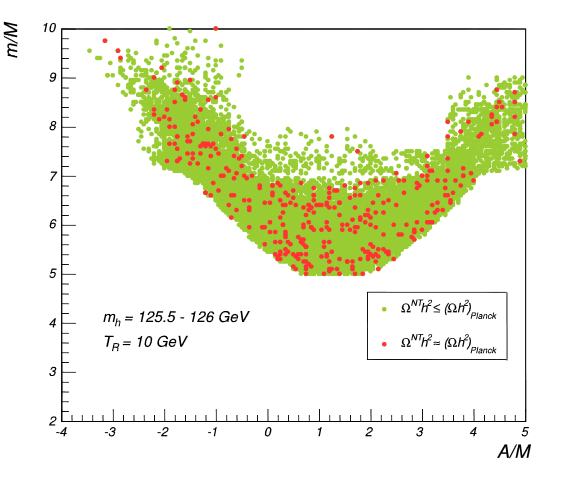}
\includegraphics[width=7.5cm]{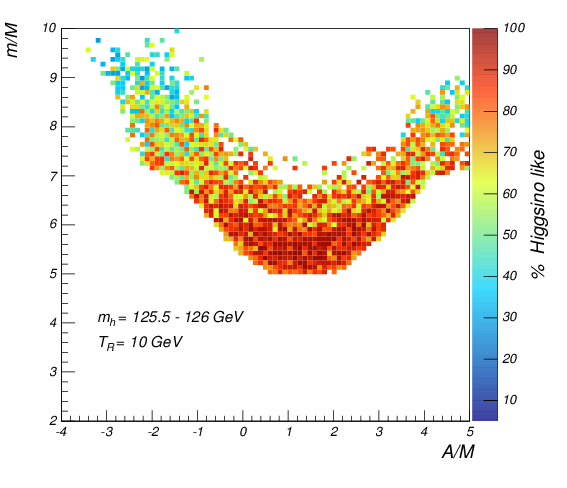}
\caption{Case with $\mu>0$: $m/M$ vs $A/M$ after imposing LEP, LHC, Planck and Fermi constraints (left)
and corresponding LSP composition (right) for $m_h = 125.5$ - $126$ GeV and $T_R = 2, 10$ GeV.}
\label{Fig6}
\end{figure}

In Fig.~\ref{Fig6} we show the results for different reheating temperatures. The red points show the parameter space where we saturate the DM content measured from Planck~\cite{planck}. We find more allowed points for $T_R=10$ GeV compared to $T_R=2$ GeV since the ratio of $T_f/T_R$ becomes smaller for large $T_R$, and so a smaller annihilation cross section is needed, resulting in a better chance to satisfy the bounds from Fermi. We only see the large Higgsino dominated regions for smaller $T_R$ since in this case a larger annihilation cross section is needed to saturate the DM content. For larger $T_R$, an LSP with a smaller Higgsino component becomes allowed (higher Bino component) and this region appears for smaller values of gaugino mass.
We finally mention that the Planck constraints on indirect detection through DM annihilation during the recombination epoch are less stringent than those of Fermi for our scenarios.

\begin{figure}[!ht]
\centering
\includegraphics[width=7.5cm]{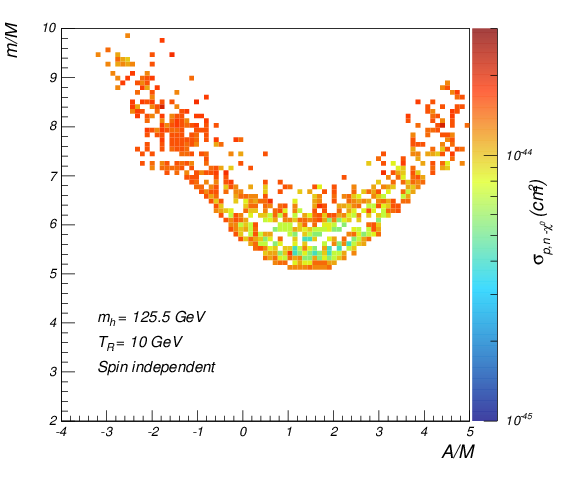}
\includegraphics[width=7.5cm]{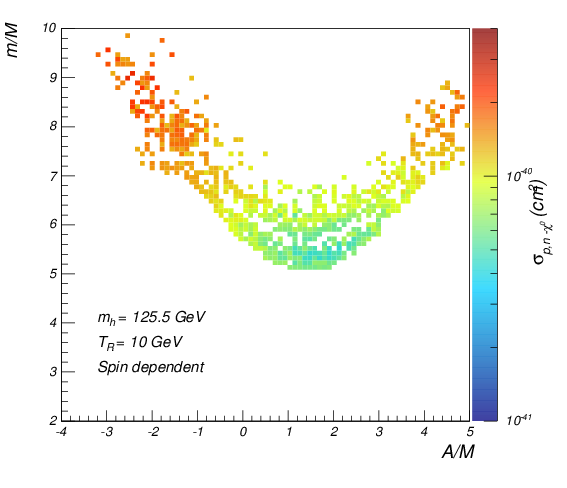}
\caption{Case with $\mu>0$: correlation between $a=m/M$ and $b=A/M$ after imposing LEP, LHC, Planck and Fermi (pass 8 limit) constraints with the corresponding spin independent (left) and spin dependent (right) WIMP-nucleon cross section for $m_h = 125.5$ GeV and $T_R = 10$ GeV.}
\label{Fig7}
\end{figure}

Fig.~\ref{Fig7} shows the spin independent and spin dependent WIMP-nucleon cross section after imposing LEP, LHC, Planck and Fermi constraints for $T_R = 10$ GeV. IceCube bounds rule out the orange and red regions of the right-hand side plot of Fig.~\ref{Fig7}.

\begin{figure}[!ht]
\centering
\includegraphics[width=7.5cm]{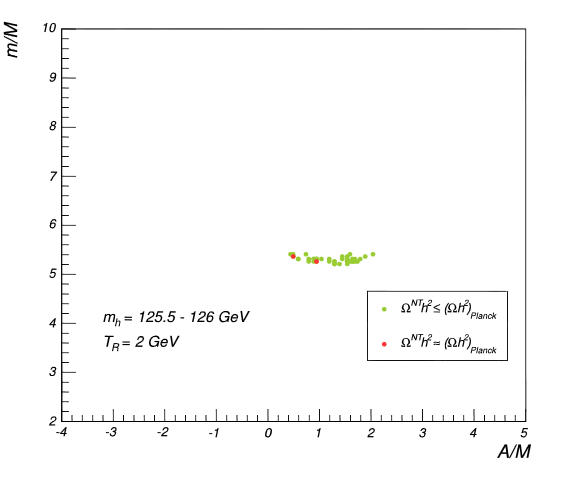}
\includegraphics[width=7.5cm]{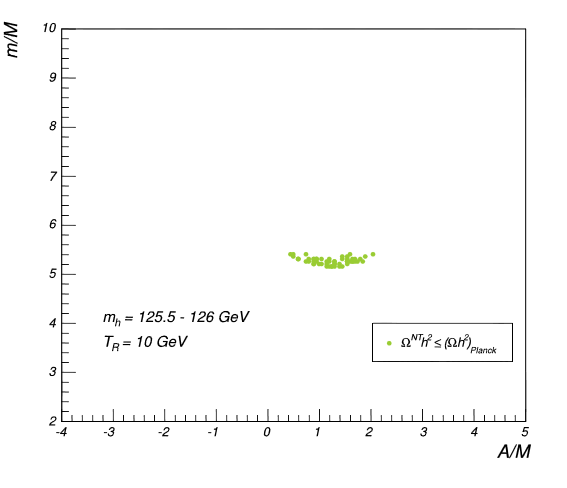}
\caption{Case with $\mu>0$: correlation between $a=m/M$ and $b=A/M$ after imposing LEP, LHC, Planck, Fermi and LUX bounds
for $m_h = 125.5$ - $126$ GeV and $T_R = 2, 10$ GeV.}
\label{Fig9}
\end{figure}

Fig.~\ref{Fig9} shows the inclusion of LUX bounds on the spin independent direct detection constraints which rule out most of the points in Fig.~\ref{Fig7} apart from a region corresponding to LSP masses around $300$ GeV which is at the border of detectability.

Moreover, for $T_R=2$ GeV there are red points which saturate the observed DM content. In this case the neutralinos are becoming more pure Higgsinos in order to enhance the annihilation cross section and the Fermi constraint becomes harder to avoid but there are still regions allowed by both direct and indirect detection searches. There are more green points for $T_R=10$ GeV since the annihilation cross section becomes smaller due to Bino mixing which means a larger allowed region after using Fermi data but the constraint from the direct detection becomes more stringent (due to Bino-Higgsino mixing in the LSP) and so there are no red points which saturate the observed DM content.\footnote{The direct detection exclusion however depends on various uncertainties, e.g.~strange quark content of proton, form factor etc.~\cite{Accomando:1999eg}.} For the points shown in Fig.~\ref{Fig9}, the GUT values $c= B/M$ and $d=\mu/M$ are around $0.6$ and $1$ respectively.

\subsubsection{Results for negative $\mu$}

\begin{figure}[!ht]
\centering
\includegraphics[width=7.5cm]{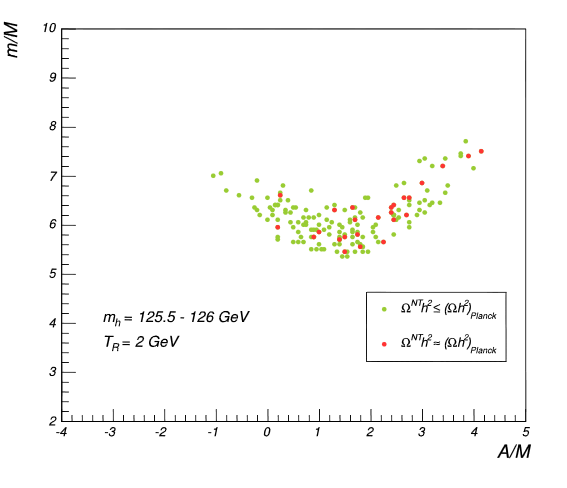}
\includegraphics[width=7.5cm]{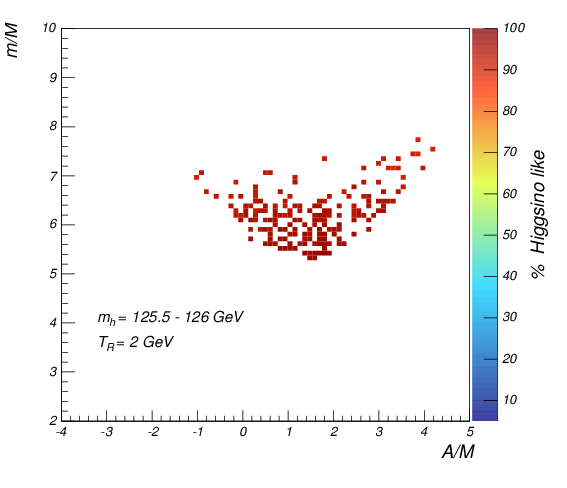}
\includegraphics[width=7.5cm]{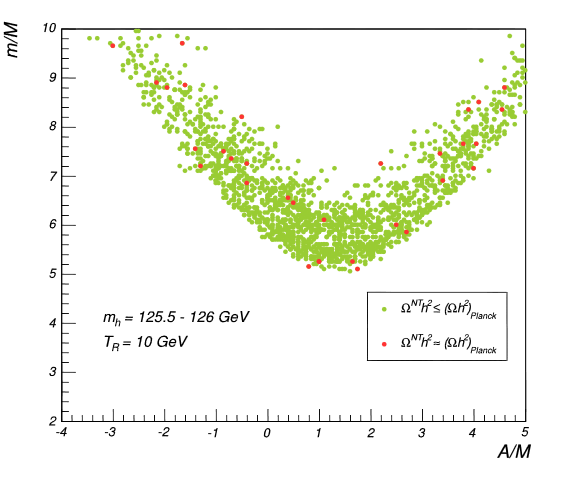}
\includegraphics[width=7.5cm]{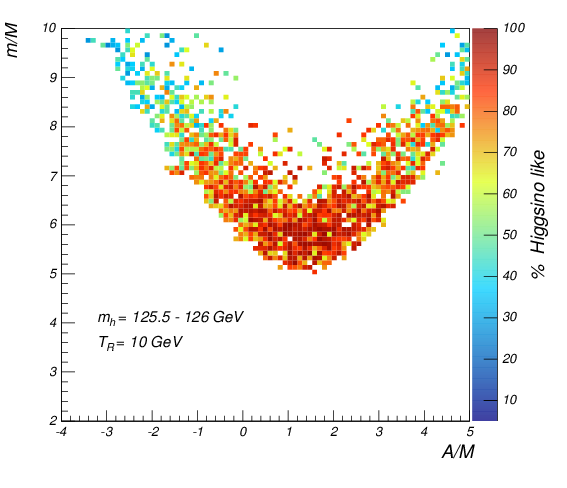}
\caption{Case with $\mu<0$: $a$ versus $b$ after imposing LEP, LHC, Planck and Fermi (pass 8 limit) bounds (left)
and LSP composition (right) for $m_h = 125.5$ - $126$ GeV and $T_R = 2, 10$ GeV.}
\label{Fig10}
\end{figure}

\begin{figure}[!ht]
\centering
\includegraphics[width=7.5cm]{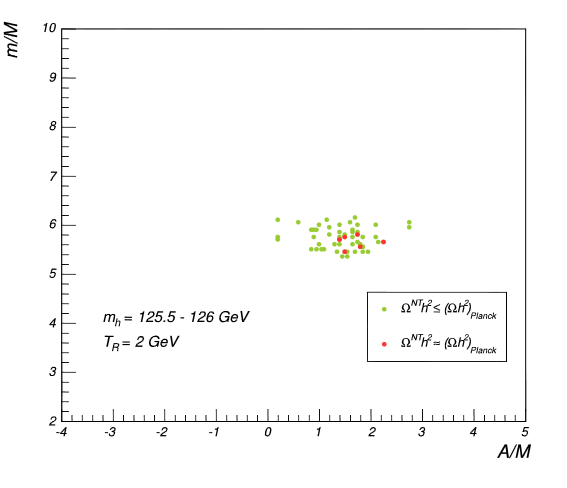}
\includegraphics[width=7.5cm]{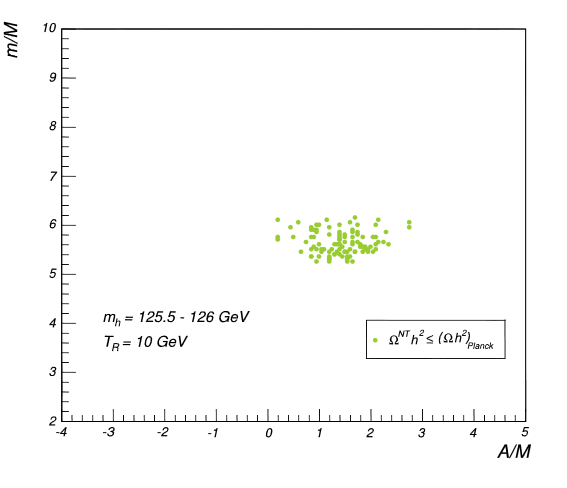}
\caption{Case with $\mu<0$: correlation between $a=m/M$ and $b=A/M$ after imposing LEP, LHC, Planck, Fermi (pass 8 limit)
and LUX for $m_h = 125.5$ - $126$ GeV and $T_R = 2, 10$ GeV.}
\label{Fig11}
\end{figure}

The results for the negative $\mu$ case are shown in Figs. \ref{Fig10} and \ref{Fig11}. 
Note that under the same conditions the allowed parameter space for $\mu<0$ is larger than the one for $\mu>0$. This can be understood as follows~\cite{positivemu}: $\sigma_{\tilde\chi^0_1-p}$ is dominated by the t-channel $h$, $H$ exchange diagrams which mostly arise from down type (s-quark) interaction:
\be
A^d\propto m_d\left({\frac{\cos\alpha}{\cos\beta}}{\frac{F_H}{m_H^2}}-{\frac{\sin\alpha}{\cos\beta}}{\frac{F_h}{m_h^2}}\right),
\ee
where $\alpha$ is the Higgs mixing angle, $F_h=(N_{12}-N_{11}\tan\theta_W)(N_{14}\cos\alpha+N_{13}\sin\alpha)$ and $F_H=(N_{12}-N_{11}\tan\theta_W)(N_{14}\sin\alpha-N_{13}\cos\alpha)$ using $\tilde\chi^0_1=N_{11}\tilde B+N_{12}\tilde W+N_{13}\tilde H_1+N_{14}\tilde H_2$. For $\mu<0$, the ratio $N_{14}/N_{13}$ is positive and this amplitude can become small due to cancellations if:
\be
\frac{N_{14}}{N_{13}}=- {\frac{\tan\alpha+m_h^2/m_H^2 \cot\alpha}{1+m_h^2/m_H^2}}\,,
\ee
is satisfied (for $\tan\alpha<0$). In Fig.~\ref{Fig11} there are more allowed points compared to Fig.~\ref{Fig9} even if there are still no points which saturate the observed DM content for $T_R=10$ GeV due to stringent direct detection constraints. For the points shown in Fig.~\ref{Fig11}, the GUT scale values of $B$ and $\mu$ are still both of order $M$.

\newpage
\section{Discussion of results}\vspace{-0.1cm}
\subsection{Analysis of the allowed parameter space}\vspace{-0.1cm}
We are now in a position to put all our results together and explain the effect of each experimental bound on our parameter space.
In the end we shall analyse the spectrum of superpartners that appears for the points surviving all the phenomenological constraints. All the observables analysed in this section have been computed numerically using \verb"micrOMEGAS" v3~\cite{micromegas} linked to \verb"SPheno" v3.3.3~\cite{spheno}.

\begin{figure}[!ht]
\centering
\includegraphics[width=7.5cm]{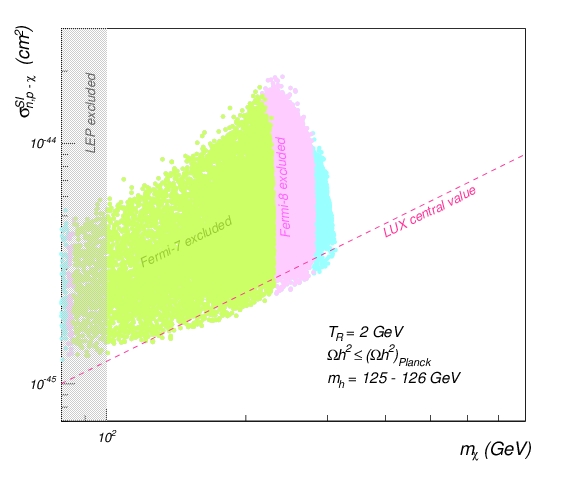}
\includegraphics[width=7.5cm]{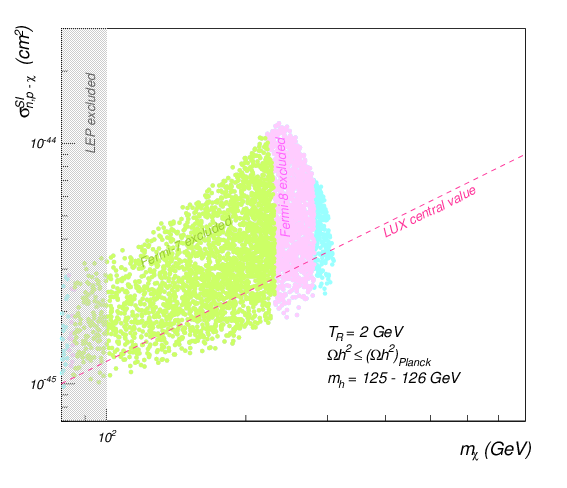}\\[-0.3cm]
\includegraphics[width=7.5cm]{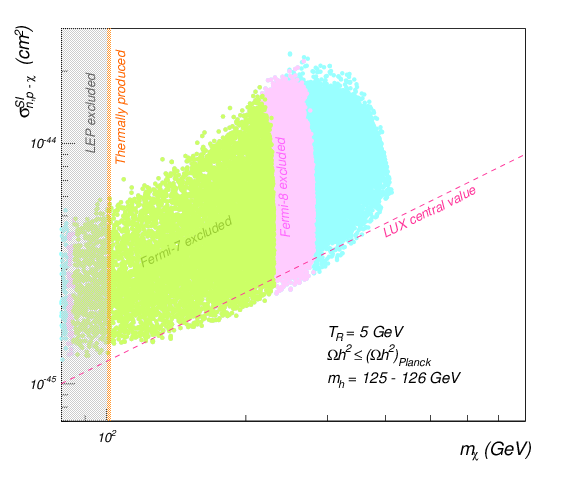}
\includegraphics[width=7.5cm]{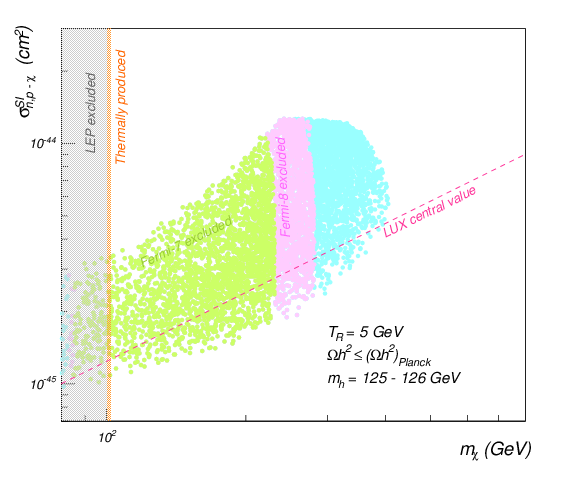}\\[-0.3cm]
\includegraphics[width=7.5cm]{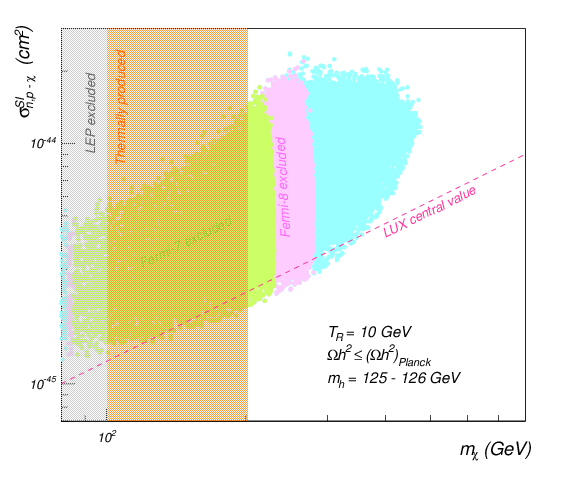}
\includegraphics[width=7.5cm]{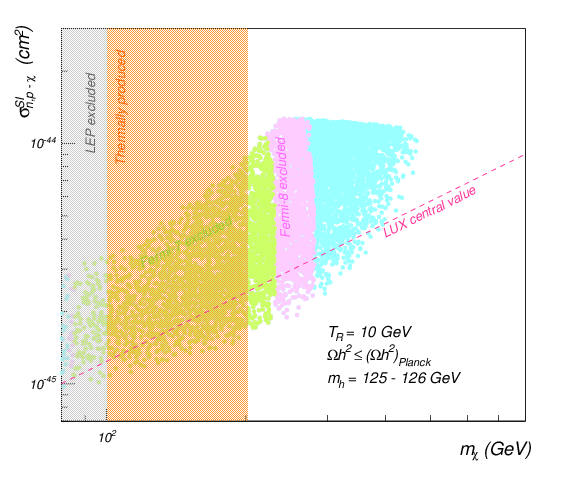}\\[-0.2cm]
\caption{Spin independent cross section versus LSP mass for $\mu>0$ (left) and $\mu<0$ (right). The light blue points are not ruled out by indirect detection experiments. We show $m_{\chi}$ up to 800 GeV.}
\label{Fig12}
\end{figure}

Fig.~\ref{Fig12} shows the relation between the spin independent WIMP-nucleon cross section and the LSP mass. Depending on $T_R$ there is a different upper bound for neutralino masses which is given by the Planck constraint on DM. For larger values of $T_R$, the non-thermal relic density decreases, 
and so heavier neutralinos can pass the Planck constraint on the DM relic density. On the other hand, for larger $T_R$ the parameter space for standard thermal DM (orange band) becomes also larger. Note that LUX 2013 results exclude at 90\% most of the parameter space and the next round of results (LUX 300 days) will be able to probe the remaining regions (the light blue points below the LUX line).

In the scenario we considered, the gaugino masses are unified at the GUT scale and therefore the evolution of electroweakinos is totally dominated by the RG flow. This implies that the LSP can only be Higgsino- or Bino-like (or a mixed combination of them). The largest contributions to the thermal averaged annihilation rates are given by (see for example~\citep{ArkaniHamed:2006mb} and references therein):
\be
\langle \sigma_{\rm eff} v \rangle = \frac{g_2^4}{512 \pi \mu^2}\left(21 + 3 \tan^2\theta_W +  11 \tan^4\theta_W   \right),
\label{sigmamu}
\ee
for Higgsino-like neutralinos (in the limit $M_W \ll \mu$) annihilating into vector bosons through chargino or neutralino interchange, and:
\be
\langle  \sigma_{\rm eff} v \rangle =  \sum_f \frac{g_2^4 \tan^2 \theta_W \left( T_{3_f} - Q_f \right)^4 r (1+r^2)}{2\pi m_{\tilde f}^2(1+r)^4}\,,
\label{sigmab}
\ee
for Bino-like LSP annihilation into fermion-antifermion ($T_{3_f}$ and $Q_f$ are the third component of isospin and the fermion charge and $r = M_1^2 / m_{\tilde f}^2$). This process is driven at tree level by the t-channel exchange of a slepton ${\tilde f}$. In the case where the LSP is a mixed composition of Higgsino and Bino, the expression of the annihilation rate is an interpolation between (\ref{sigmamu}) and (\ref{sigmab}).
Fig.~\ref{Fig12} shows also the effect of Fermi bounds. As suggested by (\ref{sigmamu}) and (\ref{sigmab}), the most constrained regions are those with smaller LSP masses. The grey band corresponds to points excluded by LEP bounds on chargino direct production.

\begin{figure}[!ht]
\centering
\includegraphics[width=7.5cm]{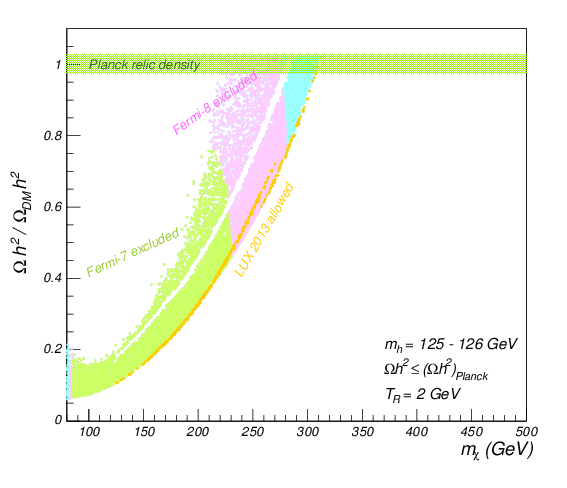}
\includegraphics[width=7.5cm]{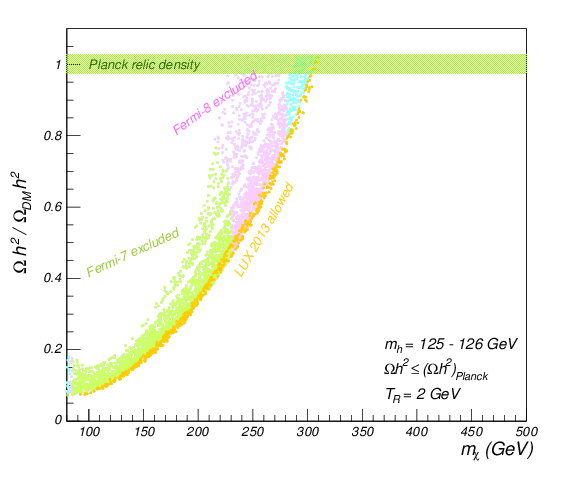}
\includegraphics[width=7.5cm]{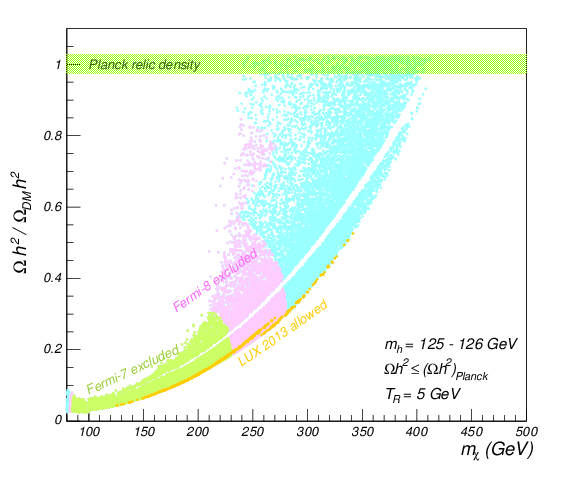}
\includegraphics[width=7.5cm]{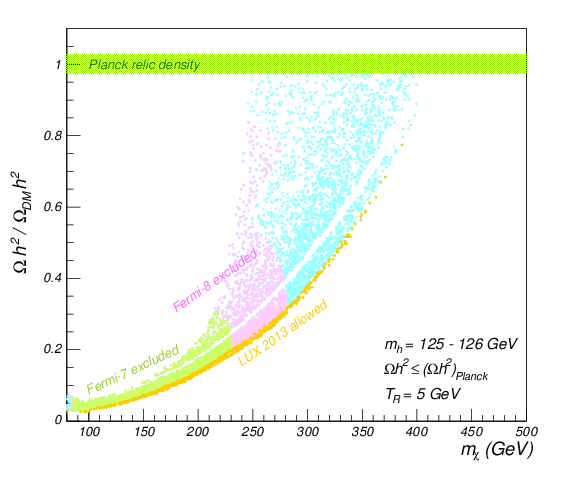}
\includegraphics[width=7.5cm]{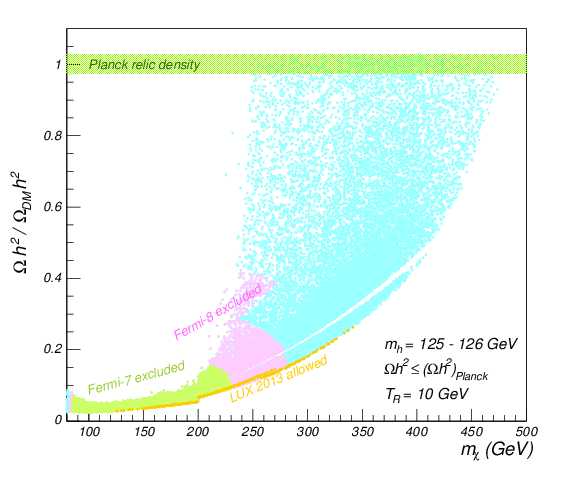}
\includegraphics[width=7.5cm]{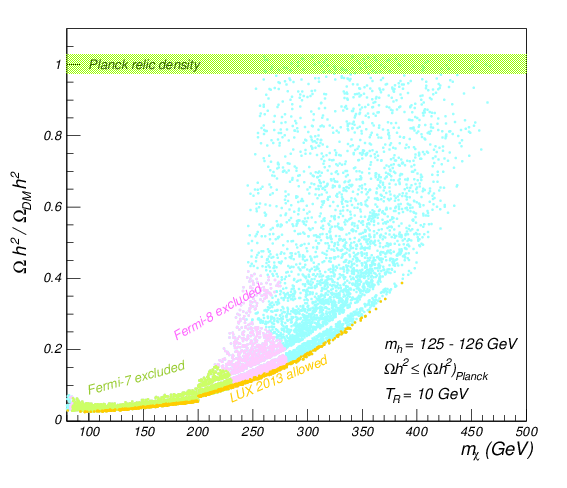}
\caption{Non-thermal DM abundance predictions versus LSP mass for $\mu>0$ (left) and $\mu<0$ (right) and $T_R=2, 5, 10$ GeV. The light blue points are not ruled out by indirect detection experiments while only the yellow points are allowed by LUX 2013 results.}
\label{Fig13}
\end{figure}

Fig.~\ref{Fig13} shows the amount of non-thermal DM relic density provided by the LSP in terms of its mass, together with the bounds from indirect detection and LUX. The Planck value of the DM content can be saturated in the region which is not ruled out by direct detection bounds only for $T_R = 2$ GeV. Given that for larger $T_R$ the amount of LSP DM gets smaller, the cases with $T_R > 2$ GeV require multi-component DM. Combining  Fig.~\ref{Fig12} and Fig.~\ref{Fig13}, we find  that the LUX allowed regions, indirect detection limits and the abundance of LSP DM are correlated for different $T_R$. The allowed regions, where the observed DM content is saturated, depend on $T_R$ but they are generically around $m_\chi \simeq 300$ GeV.

\FloatBarrier

\begin{figure}[!ht]
\centering
\includegraphics[width=7.5cm]{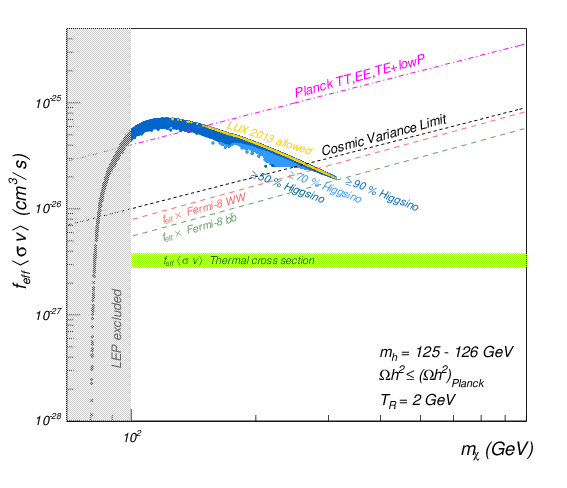}
\includegraphics[width=7.5cm]{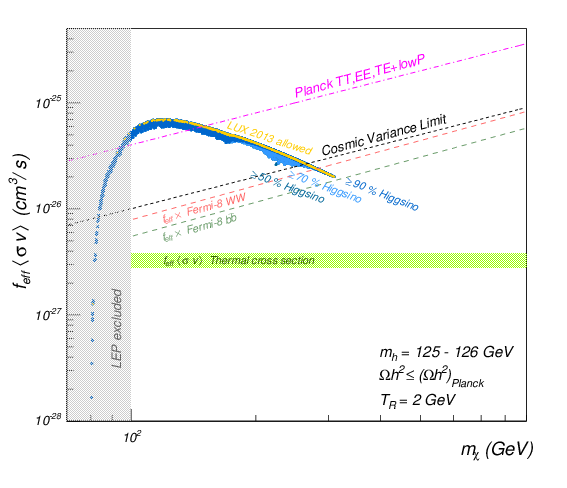}
\includegraphics[width=7.5cm]{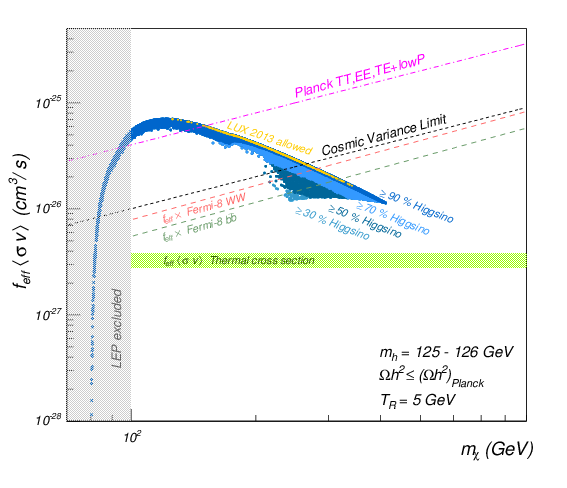}
\includegraphics[width=7.5cm]{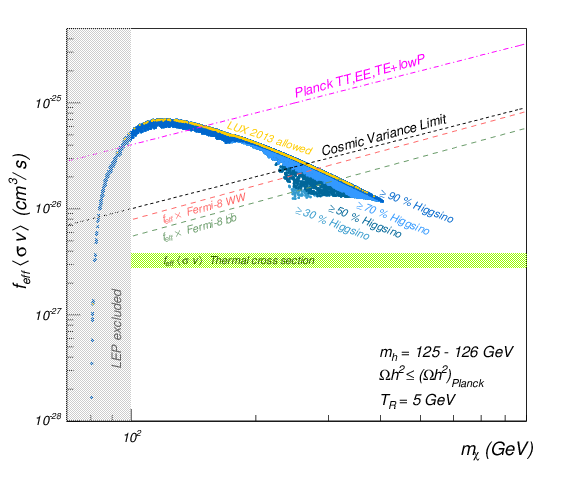}
\includegraphics[width=7.5cm]{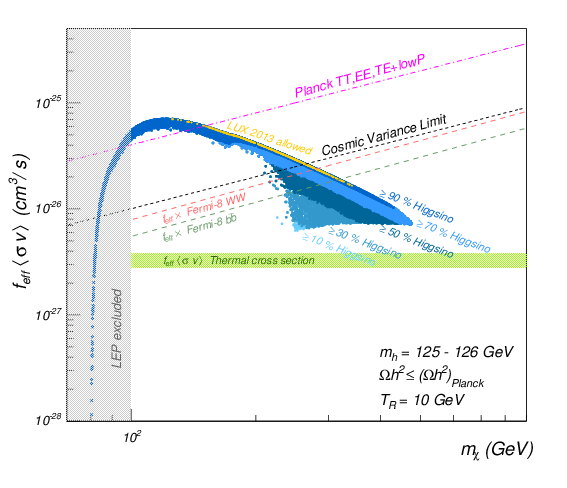}
\includegraphics[width=7.5cm]{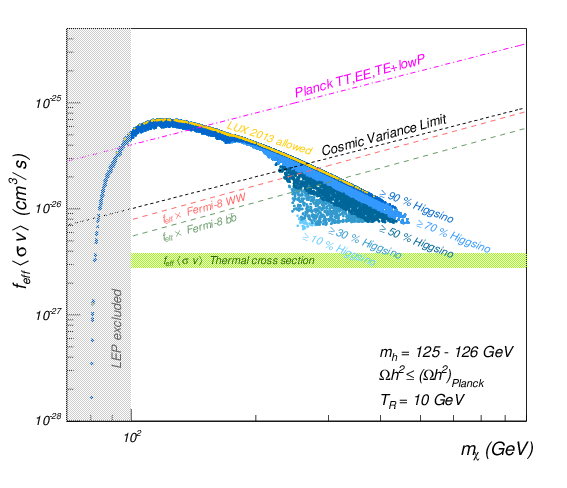}
\caption{Comparison between detection constraints from Planck, Fermi and LUX for $\mu>0$ (left) and $\mu<0$ (right). We have set $f_{\rm eff}= 0.2$. We show $m_{\chi}$ up to 900 GeV.}
\label{Fig14}
\end{figure}

In Fig.~\ref{Fig14}, we show the Planck constraints on indirect detection through DM annihilation during the recombination epoch. We use $WW$ final states corresponding to an efficiency factor $f_{\rm eff}= 0.2$~\cite{Madhavacheril:2013cna}. Even going all the way to the cosmic variance bound, these constraints turn out to be less stringent than those coming from Fermi and LUX. On the other hand, Fig.~\ref{Fig14} shows the correlation between the LSP composition and the bounds coming from both direct and indirect detection. Concerning Fermi and Planck limits on DM annihilation, these bounds allow almost all possible combinations of Bino/Higgsino neutralinos. The restrictions coming from the $f \bar f$ and $W\bar W$ channels depend on the neutralino composition: for a Higgsino-like LSP, the most stringent constraint is due to annihilation into vector bosons, while for a Bino-like LSP the main constraint comes from annihilation into a fermion-antifermion pair.

The LUX constraints in Fig.~\ref{Fig14} reduce the parameter space to the region where the LSP is mostly Higgsino-like. This could be a bit puzzling since the WIMP-nucleon cross section is dominated by the Higgs exchange channel:
\be
\sigma_{\chi-p} \propto \frac{a_{\tilde H}^2 (g' a_{\tilde B} - g \ a_{\tilde W})^2}{m_h^4}\,,
\label{xip}
\ee
where $a_{\tilde H}$, $a_{\tilde B}$ and $a_{\tilde W}$ are respectively the Higgsino, Bino and Wino LSP components. According to this expression, the cross section is enhanced when the Higgsino component increases. However in Fig.~\ref{Fig14} direct detection bounds allow only points which are mainly Higgsino-like. The reason of this effect is in the effective coupling $\tilde{\chi} \tilde{\chi}  h$ which for a Bino-like LSP looks like:
\be
C_{\tilde{\chi}\tilde{\chi} h} \simeq \frac{m_Z \sin \theta_W \tan \theta_W}{M_1^2 -\mu^2}\left( M_1 + \mu \sin 2\beta \right),
\label{Cb}
\ee
where for moderate to large $\tan \beta$ the second term is negligible and $\mu > M_1$. Hence this coupling is dominated by $M_1$. On the other hand, the coupling for a Higgsino-like LSP is:
\be
C_{\tilde{\chi}\tilde{\chi} h} \simeq \frac{1}{2}\left( 1 \pm \sin 2\beta  \right)\left( \tan^2 \theta_W\frac{m_Z \cos \theta}{M_1 - |\mu|} +\frac{m_Z \cos \theta}{M_2 - |\mu|} \right),
\label{Cn}
\ee
where $\pm$ is for the $H_u$ and $H_d$ components and $\mu < M_1$. Contrary to the Bino-like case, this coupling is inversely proportional to $M_1$. 
Thus the WIMP-nucleon cross section grows in the regions where the LSP is Higgsino-like and $M_1$ is small or where the LSP is Bino-like and the gaugino mass is large. If we compare Fig.~\ref{Fig4} (right) which shows the distribution of gaugino masses along the V-shaped band, with Fig.~\ref{Fig6} (right) and \ref{Fig10} (right), we realise that the region with a smaller cross section is the one at the bottom of the V-shaped band where gaugino masses are big and the LSP is very Higgsino-like. The region where the LSP is more Bino-like has smaller gaugino masses and the cross section is larger.
Fermi constraints however become more stringent in the case with more Higgsino content due to larger annihilation cross section. The competition between LUX and Fermi constraints produces the allowed parameter space where the Planck value of the DM content is saturated for $T_R=2$ GeV.

\FloatBarrier

\begin{figure}[!ht]
\centering
\includegraphics[width=7.5cm]{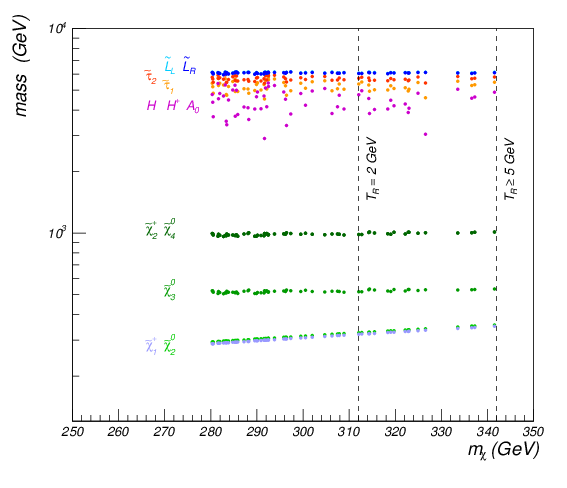}
\includegraphics[width=7.5cm]{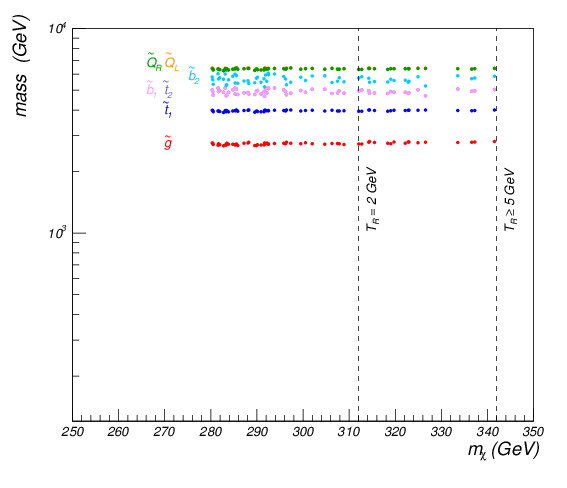}
\caption{The mass spectra of superpartners for allowed points shown in Fig.~\ref{Fig9} for different values of $T_R$.}
\label{Fig7a}
\end{figure}

Finally, in Fig.~\ref{Fig14} there is a change of behaviour of the thermal averaged cross section for masses around $130$ GeV. The reason is the following: this region is (as can be shown in the plot) Higgsino-like, but the masses are closer to $M_W$ and  $\langle \sigma v \rangle$ is no longer described by (\ref{sigmamu}) but by something like (with $x=\mu/m_W$):
\be
\langle \sigma_{\rm eff} v \rangle \sim \frac{9g^4}{16 \pi m_W^2} \frac{x^2}{(4x^2-1)^2}\,.
\ee
In Fig.~\ref{Fig7a}, we show the spectra of SUSY particles for the allowed regions of Fig.~\ref{Fig9} (blue points below the LUX line). We find that sleptons, staus, Higgses, all other scalar masses and gluinos are rather heavy since they are between about $2$ and $7$ TeV. The lightest and second to lightest neutralino and the lightest chargino are around $280$ - $340$ GeV while all other neutralinos and charginos are heavy. The allowed region for $T_R=2$ GeV is shown on the left side of the vertical line with the label $T_R=2$ GeV where the points situated exactly on the line satisfy all the constraints including the current DM content as measured by Planck. Similarly, the allowed region for $T_R\geq 5$ GeV is shown on the left side of the vertical line with the label $T_R\geq 5$ GeV even if there are no points in this region which saturate the current DM content. Notice that the spectrum is essentially independent of the reheating temperature $T_R$ and the hierarchy between the different sparticles is robust.

\FloatBarrier

\subsection{Astrophysical uncertainties}

The direct detection cross section can involve various uncertainties, e.g.~strange quark content of proton, form factor, local DM density and LSP contribution to the total amount of observed DM abundance.  The local density can be 0.1-0.7 GeV/cm$^3$~\cite{Read}. There could also be astrophysical uncertainties in the indirect detection results beyond what has been considered so far. Recently, it is  mentioned in \cite{baer} that if the thermal neutralinos do not produce the entire amount of cold dark matter, the  
 direct and indirect detection cross sections should be reduced by $R$ and $R^2$ respectively with $R\equiv \Omega h^2/0.12$. Possible bounds arising from Fermi are now almost negligible since they are suppressed by $R^2$. Once the suppression factor $R$ is taken into account, Fermi, Planck and other indirect detection experiments have lower impacts. Concerning the effect on LUX and other direct detection bounds, the cross section is now reduced by $R$ which is equivalent to multiplying the effective couplings (\ref{Cb}) and (\ref{Cn}) by $\sqrt{R}$. This clearly introduces a new parameter in 
the discussion performed in the previous section.

\begin{figure}[!ht]
\centering
\includegraphics[width=7.5cm]{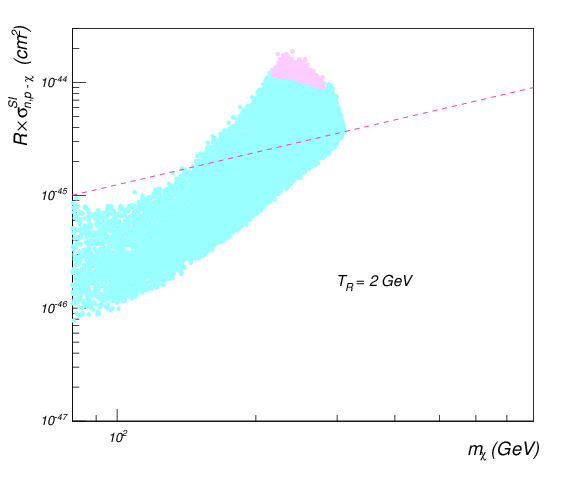}
\includegraphics[width=7.5cm]{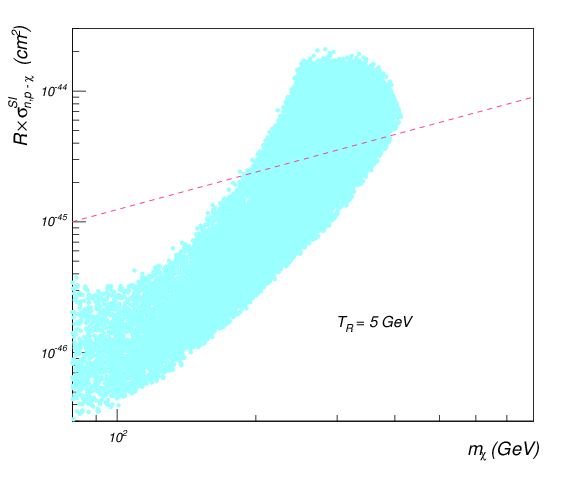}
\caption{$R\times\sigma^{SI}_{n,p-\tilde\chi^0_1}$ vs $m_{\tilde\chi^0_1}$ for $T_R=2, 5$ GeV. $R$ is defined in the text. More parameter space is allowed.}
\label{Fig15}
\end{figure}

If we assume such a reduction in the cases where $\Omega^{NT} h^2\le 0.12$, more parameter space could be allowed for multi-component DM regions as shown in Fig.~\ref{Fig15}. The pink region is disallowed by Fermi data. The $T_R$ dependence of the region constrained by Fermi in  Fig.~\ref{Fig15} is due to the fact that the  factor R is now a function of $T_R$
\begin{equation}
R = \frac{\Omega^{NT} h^2}{0.12} \simeq \frac{T_f}{T_R} \frac{\Omega^{Th} h^2}{0.12}  .
\end{equation}
$R$ becomes larger for smaller values of $T_R$ (=2 GeV) and the Fermi constraint becomes important.

The region below the dashed line satisfied present LUX limits. In particular this implies that the region with lighter neutralinos is now unconstrained by LUX. This region typically corresponds to more Bino component in the LSP as shown in Fig.~\ref{Fig6} (right) and \ref{Fig10} (right). We have therefore a different situation compared to before, because now neutralinos with a larger Bino component are allowed. 

Let us stress, however, that the prediction for the region where $\Omega^{NT} h^2$ saturates the DM content remains unchanged, i.e.~only the case $T_R=2$ GeV contains points which are still allowed by all data and saturate the DM content with an LSP mass around $300$ GeV.

This new factor $R$ helps us to extract more parameter space for the multi-component DM scenarios. However, the DM simulations need to establish the validity of  the assumption that proportions of various DM components in the early universe is maintained even after the large scale structures are formed.

\FloatBarrier

\section{Conclusions}

Non-thermal DM scenarios emerge in UV theories like string theory due to the presence of gravitationally coupled scalars which decay at late times when they are dominating the energy density of the universe. In such models the reheating temperature due to moduli decays is typically below the freeze-out temperature, $T_R<T_f$ when assuming an MSSM particle as main DM constituent. In this paper, we have studied the non-thermal version of the CMSSM/mSUGRA and contrasted it with both particle physics and astrophysical experimental constraints. The experimental information available at this moment, including the well known value of the Higgs mass, is enough to rule out the vast majority of the non-thermal CMSSM parameter space. There is still a small region which is consistent with all observations and is at the edge of detection by both astrophysics and particle physics experiments, resulting in a very interesting situation for beyond the SM physics.

In our determination of the allowed parameter space we have used constraints from collider experiments (LEP, LHC), cosmic microwave background observations (Planck) and direct and indirect DM searches (Fermi, XENON100, IceCube, LUX, CDMS). We have found that the most restrictive constraints come from Fermi and LUX which single out a small region of the non-thermal CMSSM parameter space corresponding to a Higgsino-like LSP with a mass around $300$ GeV that can saturate the observed DM abundance for $T_R\simeq 2$~GeV while larger reheating temperatures require additional contributions to the present DM abundance.
These results are summarised in Fig.~\ref{Fig13} which shows the comparison between the cases of positive and negative $\mu$.

This non-thermal scenario leads to a clear pattern of SUSY particles. In particular, the fact that the LSP is Higgsino-like makes the lightest chargino, the lightest neutralino and next to lightest neutralino to be almost degenerate in mass. This kind of scenario can be probed at the LHC using monojet plus soft leptons plus missing energy~\cite{tata}, monojet signal~\cite{mono} and two Vector Boson Fusion jets and large missing transverse energy~\cite{dutta}. On the other hand, all the other superpartners are much heavier and beyond the LHC reach but accessible to potential future experiments such as a $100$ TeV machine.

It is worth mentioning that non-thermal CMSSM scenarios with TeV-scale soft terms and reheating temperatures around $1$ - $10$ GeV can emerge in string models where the visible sector is sequestered from the sources of SUSY breaking~\cite{seqLVS, Aparicio:2014wxa}. In a subset of the parameter range these string scenarios lead to $M_{\rm soft}\sim M_P \epsilon^2 \ll m_{\rm mod}\sim M_P \epsilon^{3/2} \ll M_{\rm GUT} \sim M_P \epsilon^{1/3}$ and $T_R\sim M_P \epsilon^{9/4}$ where $\epsilon \simeq m_{3/2}/M_P\ll 1$. For $\epsilon \sim 10^{-8}$, one obtains TeV-scale soft terms, $M_{\rm GUT}\sim 10^{16}$ GeV, $T_R\sim 1 -10$ GeV and $m_{\rm mod}\sim 10^6$ GeV for $m_{3/2}\sim 10^{10}$ GeV.

We point out that our analysis is based on the CMSSM/mSUGRA for which all superpartners are expected to be at similar masses close to the TeV scale. In this sense we restricted ourselves to scalar masses lighter than $5$ TeV which is the range of validity of the codes we have used to perform our analysis. There are however several ways to generalise this simplest scenario:
\begin{itemize}
\item Consider non-universal extensions of the CMSSM. Small departures from universality, even though strongly constrained by flavour changing neutral currents, allow more flexibility in the parameter space and will slightly enhance the allowed region.

\item Consider sfermions heavier than $5$ TeV as in the split SUSY case. This is not only an interesting phenomenological possibility but is also the other class of scenarios that were derived in the string compactifications studied in~\cite{seqLVS, Aparicio:2014wxa}.

\item Consider MSSM scenarios with R-parity violation and late decaying moduli fields. This class of models has not been studied in detail and given the fact that even with R-parity conservation there seems to be a need for other sources of DM such as axions or axinos, this should be a possibility worth studying.

\item Consider explicit D-brane models which tend to generate models beyond the standard MSSM (see for instance~\cite{dolan,maharanareview} for such models).
\end{itemize}

It is encouraging that new planned experiments such as upcoming LUX result and XENON1T will be enough to rule out the rest of the allowed parameter space, independent of the upcoming LHC run. Clearly also new LHC runs and future planned colliders will be crucial for this class of models.
Combining astrophysical and collider measurements is probably the most efficient way to constrain beyond the SM physics and this article is a clear illustration of this strategy.

\section*{Acknowledgements}

We thank Shehu AbdusSalam, Giovanni Villadoro, Kerim Suruliz, Luis Ibañez, Basudeb Dasgupta, Carlos Yaguna, David G. Cerdeño, Miguel Peiró, Javier Pardo, Werner Porod, Sven Heinemayer and the ICTP ATLAS working group  for useful discussions.
MC, SK, AM and FM would like to thank the ICTP for hospitality.
The work of BD is supported in part by DOE Grant No. DE-FG02-13ER42020. The work of SK is supported by the ERC Starting Grant `Supersymmetry Breaking in String Theory.' The work of AM is supported in part by a Ramanujan fellowship.

\end{document}